\documentclass{article}
\pdfoutput=1
\usepackage{spconf,amsmath,epsfig}
\usepackage{amsmath,amssymb,amsfonts}
\usepackage[english]{babel}
\usepackage{subfig}
\usepackage{graphicx}
\usepackage{multirow}

\let\OLDthebibliography\thebibliography
\renewcommand\thebibliography[1]{
  \OLDthebibliography{#1}
  \setlength{\parskip}{0pt}
  \setlength{\itemsep}{0pt plus 0.3ex}
}

\begin{document}\sloppy

% Example definitions.
% --------------------
\def\x{{\mathbf x}}
\def\L{{\cal L}}

% Title.
% ------
\title{AUDIO CAPTIONING USING GATED RECURRENT UNITS}
%
% Single address.
% ---------------
 \name{Ay\c{s}eg\"{u}l \"{O}zkaya Eren, Mustafa Sert
	%                       \thanks{This work was supported by...}}
	               \address{Department of Computer Engineering, Ba\c{s}kent University \\
	                         21610279@mail.baskent.edu.tr, msert@baskent.edu.tr}}

\maketitle

\begin{abstract}
Audio captioning is a recently proposed task for automatically generating a textual description of a given audio clip. In this study, a novel deep network architecture with audio embeddings is presented to predict audio captions. Within the aim of extracting audio features in addition to log Mel energies, VGGish audio embedding model is used to explore the usability of audio embeddings in the audio captioning task. The proposed architecture encodes audio and text input modalities separately and combines them before the decoding stage. Audio encoding is conducted through Bi-directional Gated Recurrent Unit (BiGRU) while GRU is used for the text encoding phase. Following this, we evaluate our model by means of the newly published audio captioning performance dataset, namely Clotho, to compare the experimental results with the literature. Our experimental results show that the proposed BiGRU-based deep model outperforms the state of the art results.
\end{abstract}
\section{Introduction}
\label{sec:intro}

Audio captioning is a newly proposed task to describe the content of an audio clip using natural language sentences \cite{DBLP:journals/corr/DrossosAV17}. The purpose of creating captions is not only finding the objects, events, or scenes in the given audio clip but also finding relations between them and generating meaningful sentences. It has great potential for real-life applications such as assisting hearing impaired people and understanding environmental sounds. Additionally, since smart audio-based and video surveillance systems use audio data, audio signal analysis is a critical research area for surveillance systems. These systems can be used for recognizing activities, detecting events, anomalies, and finding semantic relations between video and audio for child-care centers, nursing homes, smart cities, elevators, etc. \cite{DBLP:journals/corr/CroccoCTM14,10.1145/3322240,8633626}. 

\begin{figure}[t]
	\centering
	
	\subfloat{%
		\raisebox{-0.5\height}{\includegraphics[width=.6\linewidth]{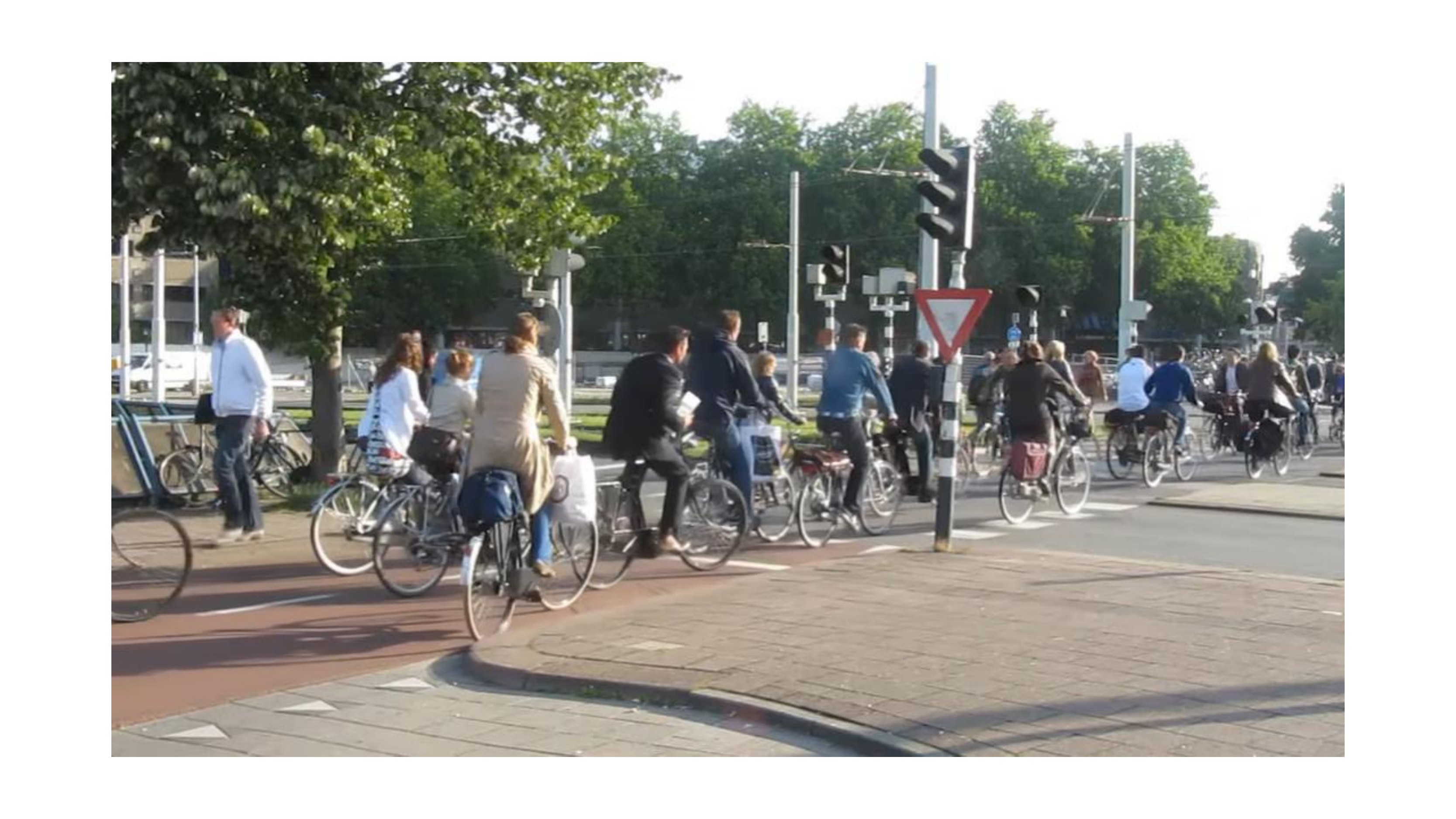}}}
	\subfloat{%
		\raisebox{-0.5\height}{\includegraphics[width=.4\linewidth]{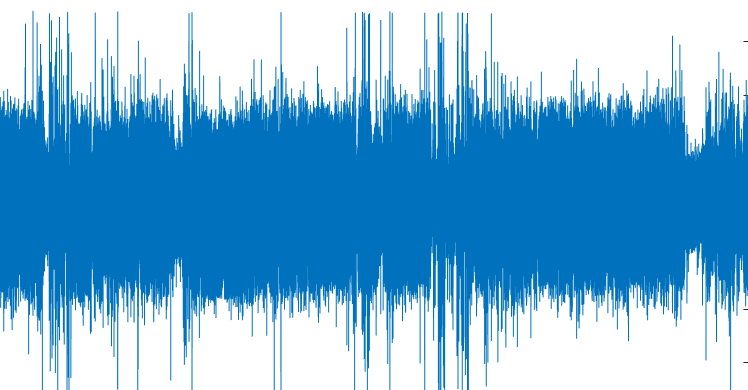}}}
\caption{A sample scene from audio-enabled video surveillance \cite{7952261}.
Without audio, there are only bicycles on the scene. With audio, bus and traffic noise are also captured.}
\end{figure}

In the field of audio signal processing, a number of tasks, such as audio event classification/detection \cite{6287923}, acoustic scene recognition \cite{7760424,8959049}, and audio tagging \cite{DBLP:journals/corr/KongXWP16} have received much attention over the past few years. In the audio event detection task, the main aim is to identify (overlapping) sound events occurring in the audio clip along with their starting and ending times. The audio tagging task assigns predefined labels to a given audio segment, whereas the acoustic scene recognition task concerns the understanding of the acoustics of the environment and assign labels to it. However, audio captioning is quite a higher level of abstraction of these tasks in the sense of generating descriptive sentences in a natural language. In audio-enabled video surveillance systems, these sentences can be used for the understanding of video scenes and possible abnormality detection within them, as well as indexing and retrieval of video (Figure 1).

Captioning is firstly used for describing images and numerous studies have been conducted \cite{DBLP:journals/corr/XuBKCCSZB15,DBLP:journals/corr/ChoCB15}. This is followed by the video captioning task, which aims to generate captions for video clips \cite{8356255,DBLP:journals/corr/abs-1804-00819}. Audio captioning task is first described in \cite{DBLP:journals/corr/DrossosAV17}. Drossos et al. propose an encoder-decoder model with three BiGRU (Bi-directional Gated Recurrent Unit) layers in the encoder and two GRU (Gated Recurrent Unit) layers in the decoder to generate audio captions by means of an attention mechanism. They use log Mel energies as audio features and a commercial dataset ProSound Effects \cite{prosound} in their experiments. Wu et al.\cite{DBLP:journals/corr/abs-1902-09254} present another attempt in the field of audio captioning. Their model is an encoder-decoder model with one GRU layer in the encoder and one GRU layer in the decoder. Also, they introduce a new audio captioning dataset for the Chinese language. An encoder-decoder model with semantic attention for generating captions for audios in the wild is presented by Kim et al. and they contribute a large scale dataset AudioCaps of 46K audio clips \cite{kim-etal-2019-audiocaps}. Drossos et al. newly introduce a publicly available audio captioning dataset called Clotho \cite{Drossos_2020} and present the results with the method in \cite{DBLP:journals/corr/DrossosAV17}.  

Our motivation is proposing a new deep network using semantic information to improve audio captioning performance. To address this problem, we propose a novel model using VGGish \cite{7952132} for audio embedding and Word2Vec \cite{DBLP:journals/corr/MikolovSCCD13} for word embedding since their performance is shown in audio classification \cite{DBLP:journals/corr/abs-1905-01926}. The core contributions of our study are as follows:

\begin{figure*}[t]
	\centering
	\subfloat[Encoding Audio Embeddings\label{subfig-1:dummy}]{%
		\includegraphics[width=.5\linewidth]{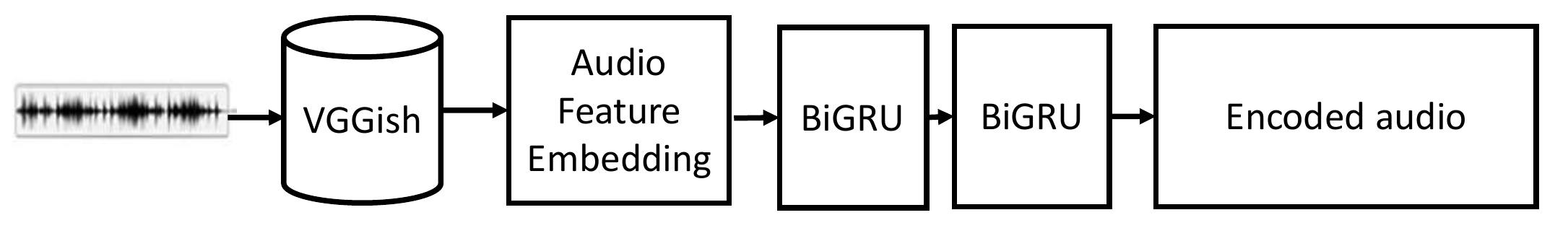}} 
	\hspace{0.5cm}
	\subfloat[Extracting Word Embeddings\label{subfig-2:dummy}]{%
		\includegraphics[width=.3\linewidth]{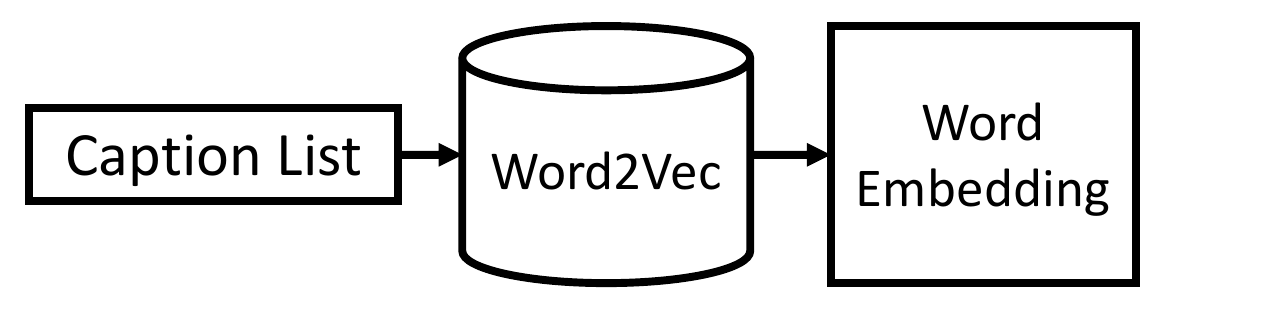}}
		
	\subfloat[Model Architecture]{\includegraphics[scale=0.5]{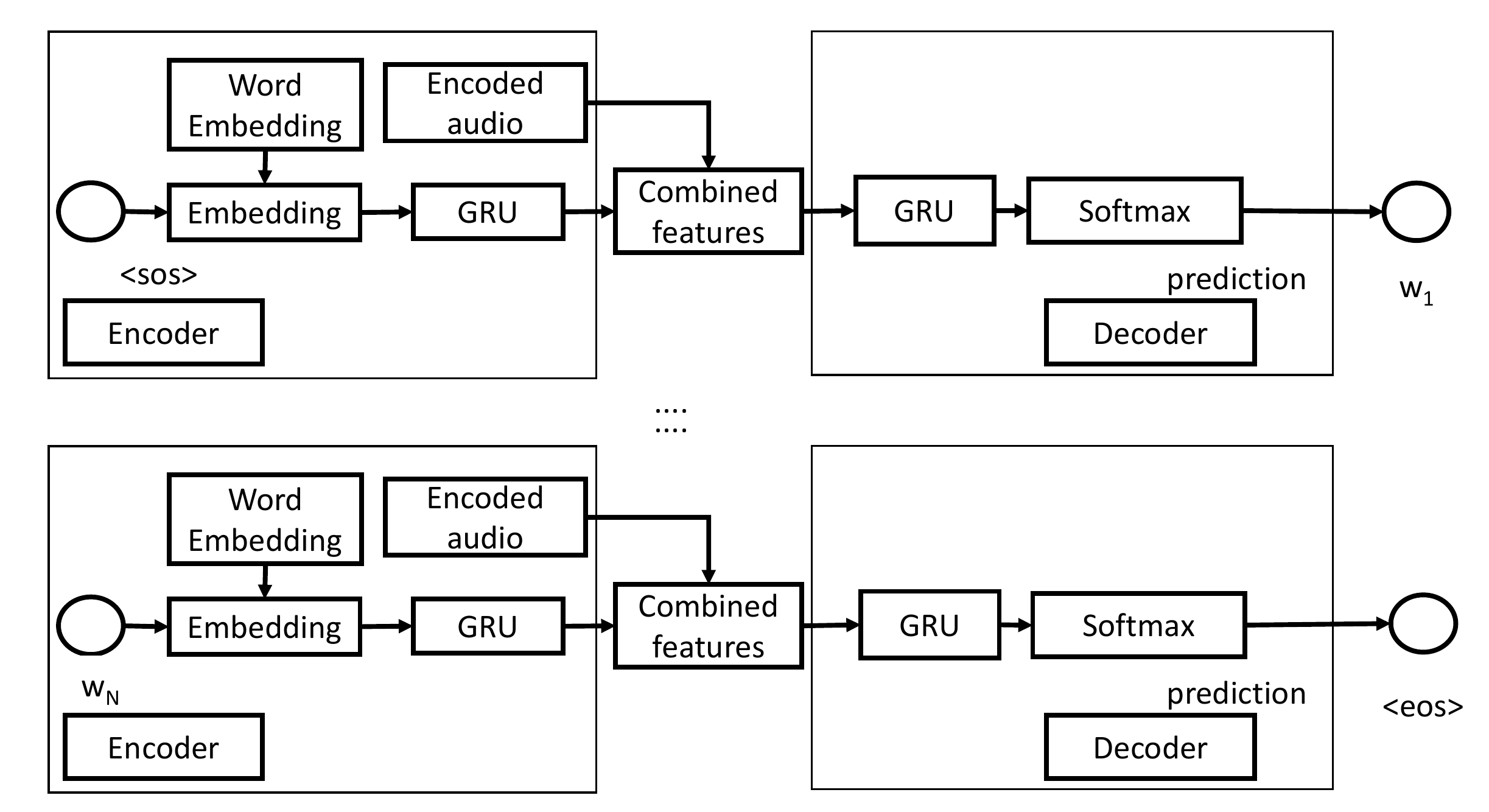}\label{fig:f2}}
	
	\caption{(a) Extracting encoded audio feature using VGGish Embeddings and BiGRUs. (b) Extraction of Word2Vec embedding to initialize weights in the embedding layer of text encoder. (c) The model merges encoded audio and encoded text, the decoder decodes given features to predict next word in the given caption. All the time frames share same weights.}\label{fig:dummy}
\end{figure*}

\begin{itemize}
	\item We propose a novel model specifically designed for the audio captioning task, which encodes audio and text separately. We combine these features and decode them in the GRU layer.
	\item The VGGish model demonstrates superior performances in the audio classification tasks \cite{7952132,DBLP:journals/corr/abs-1905-01926}. We conduct extensive experiments to demonstrate the performance of these models in the audio captioning task.	
	\item The usability of the word embedding in the audio captioning task, a well-known word embedding model Word2Vec is explored.
	\item Our model performs state-of-the-art results on the newly published audio captioning dataset Clotho. 
\end{itemize}
The organization of the paper is as follows. Section 2 introduces our proposed method. We present our experimental results and evaluations in Section 3. Finally, we give concluding remarks and possible future directions in Section 4.

\section{Proposed Method}

Our main aim is to generate meaningful captions for a given audio clip. Mathematically: 
\begin{equation}
\begin{aligned}
\theta^{\star}=\underset{\theta}{\operatorname{argmax}} 
\sum_{A,\textbf{c}} logp(\textbf{c}|A;\theta) 
\end{aligned}
\end{equation} 

We aim to maximize the probability of the caption $c$ for a given audio clip $A$ according to model parameters $\theta$. Since captions are vectors of words, $\textbf{c}$ refers to the caption of the given audio record.

\begin{equation}
\begin{aligned}
logp(\textbf{c}|A) = \sum_{t=0}^{N} logp(c_t|A,c_0,...,c_{t-1}) 
\end{aligned}
\end{equation} 

where, $N$ is the length of the caption and $c_0$ to $c_{t-1}$ is the words in the given caption.

The overall structure of our proposed model is given in Figure 2. The overall architecture consists of three modules: The audio embedding extractor, the Word2Vec word embedding extractor, and sequence modeling, which is based on RNN-GRU encoder-decoder (RNN-GRU-EncDec). The details of these modules are described in the following sections.

\subsection{Audio Feature Embedding}
We use the VGGish model to extract audio features. VGGish model is pre-trained on the AudioSet \cite{7952261}. The AudioSet is a large-scale audio event dataset and contains 2,084,320 human-labeled 10-second sound clips representing 632 audio event classes.

Previous studies show that VGGish embeddings achieve good results compared with hand-crafted audio features in audio classification tasks \cite{DBLP:journals/corr/abs-1905-01926,Basbug2019}. In order to extract audio embedding, we first extract log Mel spectrograms from audio clips. The length of the clips varies between 15 to 30 seconds. Since the length of the longest audio record is 30 seconds, we apply zero-padding to the audio records which are shorter than 30 seconds. We resample them to 16 Khz. We choose window-size of 96 milliseconds (ms) with 50\% overlap. We set the number of Mel filters to 64 similar to \cite{Drossos_2020} and frequency band to 125-7500 Hz. VGGish model extracts 128-dimensional feature vector for each second. After applying VGGish model, we obtain audio features denoted as $X=[x_1,...,x_T], x_t \in \mathbb{R}^{128}$ , where $x_t$ is a vector that contains 128 features of the audio clip and $T$ is the number of audio frames according to 96 ms window-sizes and 50\% overlaps.

\subsection{Word Embedding}

We extract word embedding using the Word2Vec model due to its superiority compared with the one-hot-encoding \cite{DBLP:journals/corr/MikolovSCCD13}. We train the Word2Vec model using the captions in Clotho development split. As a result, we generate $E=[e_1,...,e_i]$  to represent each word vector in the dataset vocabulary, where $e_i\in \mathbb{R}^{256}$, 256 is the feature dimension of word embeddings for each word. We use this pre-trained embedding to initialize weights in the embedding layer of our model. It is not used in the testing phase. 

\subsection{Encoder}

The encoding stage consists of two parts: encoding audio and encoding text. We use GRU to learn dependencies between audio frames in a given audio clip and sequences of words in captions since it reduces the number of parameters in the model \cite{DBLP:journals/corr/Lipton15}. The GRU reads whole sequence and produces one output. A simple GRU model is given as:

\begin{equation}
\begin{aligned}
z_t = \sigma(W_z.([h_{t-1},x_t]))
\end{aligned}
\end{equation} 

\begin{equation}
\begin{aligned}
r_t = \sigma(W_r.([h_{t-1},x_t]))
\end{aligned}
\end{equation} 

\begin{equation}
\begin{aligned}
\hat{h}_t = tanh(W.([r_t * h_{t-1},x_t]))
\end{aligned}
\end{equation} 

\begin{equation}
\begin{aligned}
h_t = (1-z_t) * h_{t-1} + z_t * \hat{h}_t
\end{aligned}
\end{equation} 
where  $z_t$ is the update gate at time step t, $x_t$ is the input for time step t. $W$ represents the weights, $\sigma$ is the sigmoid function, and $h_t$ is the hidden state in time step t.

Unlike feed-forward GRU, BiGRU is able to capture information not only from the past and the current state but the sequence is also reversed in time. Since an audio clip is composed as temporal sequences of frames, we use BiGRU to learn the relationship between audio time steps. We use two BiGRU layers in our design. In the encoding stage of our model, the first BiGRU layer has 32 cells and second has 64 cells, which are selected empirically. For text encoding, Word2Vec model weights are used to initialize our model’s word embedding layer. This embedding is given to the first GRU layer which has 128 cells. This GRU is used to learn word sequences. In order to combine encoded audio and text, we use the addition method.

\subsection{Decoder}

We design the decoder with a single GRU layer consisting of 128 cells. Its inputs combined feature vector from the encoder and outputs the next predicted word. We use the $Softmax$ after the fully connected layer. The decoder performs the prediction word by word and a sequence of the predicted words gives the caption.

\section{EXPERIMENTS AND RESULTS}
\subsection{Dataset}

We conduct our experiments on the newly published dataset Clotho \cite{Drossos_2020}. The development and the evaluation sets of the dataset contain 2893 and 1043 audio clips, respectively. Both of the sets have 5 captions for each audio clip. The lengths of the audio clips is 15 to 30 seconds in duration and captions are 8 to 20 words. We split the development dataset into two parts which are training and validation. 2000 audio records are selected randomly for training and the remaining part is used for validation. We use each audio clip five times with one assigned caption from the caption-list based on the best practice in \cite{Drossos_2020}. For instance, let $a_i$ is an individual audio clip with captions  $S={[s_1,s_2,..,s_5]}$, then we use this audio clip instance as 5 separate instances: $<a_i,s_1>, <a_i,s_2>, .., <a_i,s_5>$ in the training. To find start and end of the sequences of captions, we add special $<sos>$ and $<eos>$ in the beginning and end of the captions.

\subsection{Training Details}

The proposed model has approximately 2,000,000 parameters. Adam optimizer and LeakyRelu activation function are used in the training. Batch-size is set to 64. We use a dropout rate of 0.5 for input connections. Batch normalization \cite{DBLP:journals/corr/IoffeS15} is used after each BiGRU and GRU layer in the encoding and decoding phases. Loss function is categorical-cross entropy since it is widely used in the literature \cite{tanti-etal-2017-role}. It is given by  

\begin{equation}
\begin{aligned}
L(\Theta) =- \sum_{t=1}^{T} log(p_\Theta(w_t|w_1,...,w_{t-1})
\end{aligned}
\end{equation} 
where $w_t$ is the predicted word based on previous words.

To prevent gradient vanishing problem, LeakyReLU activation function is chosen empirically, where the LeakyReLU is given by, 

\begin{equation}
\begin{aligned}
lrelu(x) =
\begin{cases}
x & \text{x$>$0}\\
\alpha & \text{x$\leq$0}\\
\end{cases} 
\end{aligned}
\end{equation}
 where $\alpha$ is chosen 0.3 in this study which is the default value of LeakyReLU in Keras \cite{keras}. It uses small gradient when the cell is not active.
 
The final hyperparameters such as the batch-size, dropout rate, and activation functions used in the study are chosen based on minimum validation loss in our several experiments. We implemented the system using Keras framework and run on a computer with GPU GTX1660Ti in a system Linux Ubuntu 18.04 and Python 3.6. The model is run for 50 epochs. In the experiments, 1 epoch with Mel-energy features takes approximately 4 hours whereas 1 epoch with the VGGish model takes approximately 15 minutes according to the given configurations. The minimum validation error is obtained in the 30th epoch for the VGGish model given in Figure 3.

 \begin{figure}[t]
 	\centering
 	\includegraphics[width=.9\linewidth]{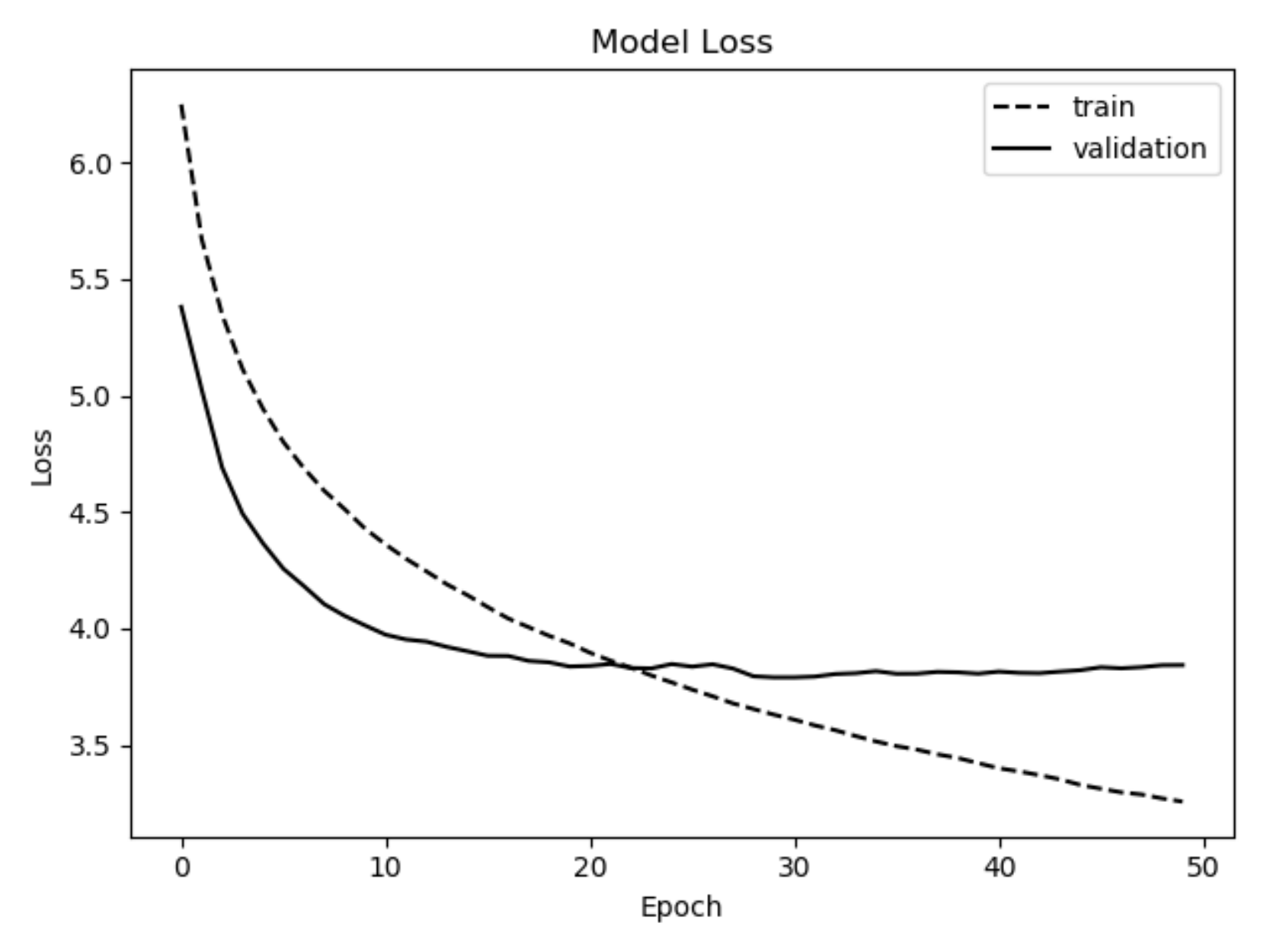}
 	\caption{The proposed method loss-validation loss plot}
 	\label{figTraining}
 \end{figure}
\subsection{Evaluation}

We perform our evaluations on the public performance dataset Clotho and compare our results with the method introduced with the Clotho [15]. We evaluate our experiments with widely used metrics in machine translation tasks and also used in the Clotho. To this aim, we use the BLEU \cite{Papineni02bleu:a}, METEOR \cite{banerjee-lavie-2005-meteor}, CIDEr \cite{DBLP:journals/corr/VedantamZP14a}, and ROUGE$L$ \cite{lin-2004-rouge} metrics for the evaluations.

\begin{table*}[t]
	
		\caption{Performance comparison of the proposed method. RNN-GRU-EncDec is the proposed encoder-decoder based architecture for sequence modeling. (BLEU-1: B-1, BLEU-2: B-2, BLEU-3: B-3, BLEU-4: B-4)}
		\begin{center}
		\begin{tabular}{ |l|l|c|c|c|c|c|c|}
			\hline
			 \multirow{2}*{\bfseries \hspace{2.5cm} Method} & \multicolumn{7} {c|}{\bfseries Metric} \\
			\cline{2-8}
		     & \textbf{B-1}& \textbf{B-2}& \textbf{B-3} & \textbf{B-4} & \textbf{CIDEr} & \textbf{METEOR} & \textbf{ROUGE$_L$}\\
			\hline
			\textbf{Clotho \cite{Drossos_2020}}    &       0.42     &   0.14 &  0.06      &    0.02   &    0.10 &    0.09 &    0.27\\
			\hline
			\textbf{RNN-GRU-EncDec + Log Mel Energy}   &       0.45    &  0.21 &   0.16      &    0.08  &    0.11 &    0.17 &    0.34 \\
			\hline 
			\textbf{RNN-GRU-EncDec + VGGish}  &       0.45    &   0.24 &  0.18      &     0.09 &    0.15  &    0.18 &    0.38\\
			\hline 
			\textbf{RNN-GRU-EncDec + VGGish + Word2Vec}   &       0.51    &  0.28 &   0.22      &     0.12  &    0.18 &    0.19 &    0.40 \\
			\hline 
		
		\end{tabular}
	    \end{center}
	\label{table2}
\end{table*}

The metric BLEU${_n}$ calculates the precision for n-grams. To calculate precision, the matching words in the actual sentence and the predicted sentence is calculated. BLEU does not consider the context of the word in the sentence. The metric range is between [0,1]. If the actual sentence and the predicted sentence is totally the same, then the score is 1. BLEU-1 (B-1) represents 1-gram, whereas BLEU-4 represents 4-grams. METEOR calculates recall and precision together and takes a harmonic mean score. It creates an alignment between actual and predicted sentences and makes mapping between them. CIDEr also uses n-gram model and it calculates cosine-similarity between the actual and predicted sentences. It also considers the Term Frequency Inverse-Document Frequency. ROUGE$_L$ calculates Longest Common Subsequences which considers the sequence of the words in the actual and predicted sentences.

\subsection{Results}

We compare our results with the method in the Clotho dataset. To show Word2Vec contribution in our model, we present our results with and without Word2Vec embedding. Our experimental results are presented in Table 1. The results show that our model outperforms the state-of-the-art. 

The results show that our proposed model with log-Mel features has better results than the literature. The proposed model with VGGish embeddings provides better results than log-Mel features. Additionally, VGGish provide better training performance in terms of time and memory usage.

The predicted sentences show that our model can generally predict the main content of the audio clip. For instance, our model predicts "People are talking and laughing" whereas the ground truth is "People are talking and laughing with loud person near the end". It predicts sentence in correct order but shorter than the ground truth. 

In our proposed model, similar concepts are also predicted. To illustrate, our model predicts "Rain is falling heavily and thunder is booming" while the ground truth is "Passing windstorm outside and something is striking against another harder object". Actually they are similar concepts but according to BLEU, it is not assessed as a successful instance because the metric is based on calculating precision on exactly the same words. As another example, our model predicts the caption as "Bicycle is coasting down road slowly" whereas the ground truth is "The engine of vehicle is driving down the road". In this example, our model does not differentiate the bicycle and engine sounds. Some other predicted captions are given below to show our model’s performance.

\noindent\textbf{Actual-1}:{\fontfamily{pcr}\selectfont Busy restaurant with people eating during rush hour}\\
\textbf{Actual-2}:{\fontfamily{pcr}\selectfont Crowd of people are walking and talking} \\
\textbf{Actual-3}:{\fontfamily{pcr}\selectfont Crowded restaurant with people eating during rush hour} \\
\textbf{Actual-4}:{\fontfamily{pcr}\selectfont Hall filled with conversing people echoes with talk} \\
\textbf{Actual-5}:{\fontfamily{pcr}\selectfont People chatting in the hall down fair distance with an echo} \\
\textbf{Prediction}:{\fontfamily{pcr}\selectfont Group of people are talking and laughing in a crowded place} \\
\\
\textbf{Actual-1}:{\fontfamily{pcr}\selectfont Person walking back and forth in the rain as car pass} \\
\textbf{Actual-2}:{\fontfamily{pcr}\selectfont Birds are singing as someone walks by thunder roars and vehicles drive past} \\
\textbf{Actual-3}:{\fontfamily{pcr}\selectfont Footsteps over dog barking while the wind blows} \\
\textbf{Actual-4}:{\fontfamily{pcr}\selectfont Person is walking back and forth in the rain as car pass by} \\
\textbf{Actual-5}:{\fontfamily{pcr}\selectfont Vehicles driving past birds singing someone walking and thunder in the distance}\\ 
\textbf{Prediction}:{\fontfamily{pcr}\selectfont Birds are chirping and singing while cars are passing}
\section{CONCLUSIONS}
In this paper, we present a novel model that combines text and audio features to predict audio captions. The proposed model uses VGGish audio embeddings that provide us semantically embedding and smaller feature dimension than raw audio features and log-Mel band energies. It brings us better training performance. Word2Vec is used to extract word embedding and results show that semantic information can improve audio captioning performance. A novel encoder-decoder model is designed with multiple BiGRU and GRU layers. We evaluate our study in a newly published public dataset called Clotho. Experiments show that our proposed method yields better performance than the state-of-the-art studies. 

The results show that our model is able to predict audio captions. The predicted captions are more general and shorter than the actual truths. The proposed model does not differentiate similar sounds. It can explicitly be stated that we can obtain better results if we have a larger dataset and train it for more epochs. Furthermore, the model generally predicts similar-structured sentences. Improving the language model and adding semantic information can increase the performance.

According to these results, our future research direction is to strive for improving language modeling and to use data augmentation techniques in an attempt of enhancing the performance of our model. Getting better results on audio captioning can yield improvement in audio analysis. Additionally, multimodal models can be researched to improve the performance of video applications such as video captioning, video retrieval, and surveillance systems which are mainly composed of audio and video analysis.

\bibliographystyle{IEEEbib}
\bibliography{icme2020template}

\end{document}